\documentclass[aps,prl,superscriptaddress,showpacs,floatfix,twocolumn]{revtex4}

\usepackage{graphicx}	

\newcommand{\pperp}{\mbox{$p_\mathrm{T}\,$}}

\def\gsim{\raise0.3ex\hbox{$>$\kern-0.75em\raise-1.1ex\hbox{$\sim$}}}

\begin{document}

\title{Centrality Dependence of $\pi^0$ and $\eta$ Production at Large Transverse
Momentum in $\sqrt{s_{\scriptscriptstyle NN}}$~=~200~GeV $d$+Au Collisions}

\newcommand{\abilene}{Abilene Christian University, Abilene, TX 79699, USA}
\newcommand{\acadsin}{Institute of Physics, Academia Sinica, Taipei 11529, Taiwan}
\newcommand{\banaras}{Department of Physics, Banaras Hindu University, Varanasi 221005, India}
\newcommand{\barc}{Bhabha Atomic Research Centre, Bombay 400 085, India}
\newcommand{\bnl}{Brookhaven National Laboratory, Upton, NY 11973-5000, USA}
\newcommand{\caucr}{University of California - Riverside, Riverside, CA 92521, USA}
\newcommand{\ciae}{China Institute of Atomic Energy (CIAE), Beijing, People's Republic of China}
\newcommand{\cns}{Center for Nuclear Study, Graduate School of Science, University of Tokyo, 7-3-1 Hongo, Bunkyo, Tokyo 113-0033, Japan}
\newcommand{\colorado}{University of Colorado, Boulder, CO 80309, USA}
\newcommand{\columbia}{Columbia University, New York, NY 10027 and Nevis Laboratories, Irvington, NY 10533, USA}
\newcommand{\dapnia}{Dapnia, CEA Saclay, F-91191, Gif-sur-Yvette, France}
\newcommand{\debrecen}{Debrecen University, H-4010 Debrecen, Egyetem t{\'e}r 1, Hungary}
\newcommand{\elte}{ELTE, E{\"o}tv{\"o}s Lor{\'a}nd University, H - 1117 Budapest, P{\'a}zm{\'a}ny P. s. 1/A, Hungary}
\newcommand{\fsu}{Florida State University, Tallahassee, FL 32306, USA}
\newcommand{\gsu}{Georgia State University, Atlanta, GA 30303, USA}
\newcommand{\hiroshima}{Hiroshima University, Kagamiyama, Higashi-Hiroshima 739-8526, Japan}
\newcommand{\ihepprot}{IHEP Protvino, State Research Center of Russian Federation, Institute for High Energy Physics, Protvino, 142281, Russia}
\newcommand{\illuiuc}{University of Illinois at Urbana-Champaign, Urbana, IL 61801, USA}
\newcommand{\isu}{Iowa State University, Ames, IA 50011, USA}
\newcommand{\jinrdubna}{Joint Institute for Nuclear Research, 141980 Dubna, Moscow Region, Russia}
\newcommand{\kek}{KEK, High Energy Accelerator Research Organization, Tsukuba, Ibaraki 305-0801, Japan}
\newcommand{\kfki}{KFKI Research Institute for Particle and Nuclear Physics of the Hungarian Academy of Sciences (MTA KFKI RMKI), H-1525 Budapest 114, POBox 49, Budapest, Hungary}
\newcommand{\korea}{Korea University, Seoul, 136-701, Korea}
\newcommand{\kurchatov}{Russian Research Center ``Kurchatov Institute", Moscow, Russia}
\newcommand{\kyoto}{Kyoto University, Kyoto 606-8502, Japan}
\newcommand{\labllr}{Laboratoire Leprince-Ringuet, Ecole Polytechnique, CNRS-IN2P3, Route de Saclay, F-91128, Palaiseau, France}
\newcommand{\lawllnl}{Lawrence Livermore National Laboratory, Livermore, CA 94550, USA}
\newcommand{\losalamos}{Los Alamos National Laboratory, Los Alamos, NM 87545, USA}
\newcommand{\lpc}{LPC, Universit{\'e} Blaise Pascal, CNRS-IN2P3, Clermont-Fd, 63177 Aubiere Cedex, France}
\newcommand{\lund}{Department of Physics, Lund University, Box 118, SE-221 00 Lund, Sweden}
\newcommand{\muenster}{Institut f\"ur Kernphysik, University of Muenster, D-48149 Muenster, Germany}
\newcommand{\myongji}{Myongji University, Yongin, Kyonggido 449-728, Korea}
\newcommand{\nagasaki}{Nagasaki Institute of Applied Science, Nagasaki-shi, Nagasaki 851-0193, Japan}
\newcommand{\newmex}{University of New Mexico, Albuquerque, NM 87131, USA }
\newcommand{\nmsu}{New Mexico State University, Las Cruces, NM 88003, USA}
\newcommand{\ornl}{Oak Ridge National Laboratory, Oak Ridge, TN 37831, USA}
\newcommand{\orsay}{IPN-Orsay, Universite Paris Sud, CNRS-IN2P3, BP1, F-91406, Orsay, France}
\newcommand{\peking}{Peking University, Beijing, People's Republic of China}
\newcommand{\pnpi}{PNPI, Petersburg Nuclear Physics Institute, Gatchina, Leningrad region, 188300, Russia}
\newcommand{\riken}{RIKEN (The Institute of Physical and Chemical Research), Wako, Saitama 351-0198, JAPAN}
\newcommand{\rikjrbrc}{RIKEN BNL Research Center, Brookhaven National Laboratory, Upton, NY 11973-5000, USA}
\newcommand{\saopaulo}{Universidade de S{\~a}o Paulo, Instituto de F\'{\i}sica, Caixa Postal 66318, S{\~a}o Paulo CEP05315-970, Brazil}
\newcommand{\seoulnat}{System Electronics Laboratory, Seoul National University, Seoul, South Korea}
\newcommand{\stonybrkc}{Chemistry Department, Stony Brook University, Stony Brook, SUNY, NY 11794-3400, USA}
\newcommand{\stonycrkp}{Department of Physics and Astronomy, Stony Brook University, SUNY, Stony Brook, NY 11794, USA}
\newcommand{\subatech}{SUBATECH (Ecole des Mines de Nantes, CNRS-IN2P3, Universit{\'e} de Nantes) BP 20722 - 44307, Nantes, France}
\newcommand{\tenn}{University of Tennessee, Knoxville, TN 37996, USA}
\newcommand{\titech}{Department of Physics, Tokyo Institute of Technology, Oh-okayama, Meguro, Tokyo 152-8551, Japan}
\newcommand{\tsukuba}{Institute of Physics, University of Tsukuba, Tsukuba, Ibaraki 305, Japan}
\newcommand{\vandy}{Vanderbilt University, Nashville, TN 37235, USA}
\newcommand{\waseda}{Waseda University, Advanced Research Institute for Science and Engineering, 17 Kikui-cho, Shinjuku-ku, Tokyo 162-0044, Japan}
\newcommand{\weizmann}{Weizmann Institute, Rehovot 76100, Israel}
\newcommand{\yonsei}{Yonsei University, IPAP, Seoul 120-749, Korea}
\newcommand{\deceased}{\dagger}
\affiliation{\abilene}
\affiliation{\acadsin}
\affiliation{\banaras}
\affiliation{\barc}
\affiliation{\bnl}
\affiliation{\caucr}
\affiliation{\ciae}
\affiliation{\cns}
\affiliation{\colorado}
\affiliation{\columbia}
\affiliation{\dapnia}
\affiliation{\debrecen}
\affiliation{\elte}
\affiliation{\fsu}
\affiliation{\gsu}
\affiliation{\hiroshima}
\affiliation{\ihepprot}
\affiliation{\illuiuc}
\affiliation{\isu}
\affiliation{\jinrdubna}
\affiliation{\kek}
\affiliation{\kfki}
\affiliation{\korea}
\affiliation{\kurchatov}
\affiliation{\kyoto}
\affiliation{\labllr}
\affiliation{\lawllnl}
\affiliation{\losalamos}
\affiliation{\lpc}
\affiliation{\lund}
\affiliation{\muenster}
\affiliation{\myongji}
\affiliation{\nagasaki}
\affiliation{\newmex}
\affiliation{\nmsu}
\affiliation{\ornl}
\affiliation{\orsay}
\affiliation{\peking}
\affiliation{\pnpi}
\affiliation{\riken}
\affiliation{\rikjrbrc}
\affiliation{\saopaulo}
\affiliation{\seoulnat}
\affiliation{\stonybrkc}
\affiliation{\stonycrkp}
\affiliation{\subatech}
\affiliation{\tenn}
\affiliation{\titech}
\affiliation{\tsukuba}
\affiliation{\vandy}
\affiliation{\waseda}
\affiliation{\weizmann}
\affiliation{\yonsei}
\author{S.S.~Adler}	\affiliation{\bnl}
\author{S.~Afanasiev}	\affiliation{\jinrdubna}
\author{C.~Aidala}	\affiliation{\columbia}
\author{N.N.~Ajitanand}	\affiliation{\stonybrkc}
\author{Y.~Akiba}	\affiliation{\kek} \affiliation{\riken}
\author{A.~Al-Jamel}	\affiliation{\nmsu}
\author{J.~Alexander}	\affiliation{\stonybrkc}
\author{K.~Aoki}	\affiliation{\kyoto}
\author{L.~Aphecetche}	\affiliation{\subatech}
\author{R.~Armendariz}	\affiliation{\nmsu}
% ...
\collaboration{PHENIX Collaboration} \noaffiliation

\date{\today}

\begin{abstract}

The dependence of transverse momentum spectra of neutral pions and $\eta$ mesons with
$p_\mathrm{T} < 16$~GeV/$c$ and $p_\mathrm{T} < 12$~GeV/$c$, respectively, on the centrality of the collision 
has been measured at 
mid-rapidity by the PHENIX experiment at RHIC in $d$+Au collisions
at $\sqrt{s_{\scriptscriptstyle NN}}$ = 200~GeV.
The measured yields are compared to those in $p+p$ collisions at the
same $\sqrt{s_{\scriptscriptstyle NN}}$ scaled by the number of underlying 
nucleon-nucleon collisions in $d$+Au. 
At all centralities the yield ratios show no suppression, in contrast
to the strong suppression seen for central Au+Au collisions at RHIC.
Only a weak \pperp and centrality dependence can be observed.

\end{abstract}

\pacs{25.75.Dw}


\maketitle


%
High-energy nucleus-nucleus collisions provide the opportunity to
study strongly interacting matter at very high energy densities
where Quantum Chromodynamics (QCD) predicts a transition from normal 
nuclear matter to a deconfined system of quarks and gluons, 
the Quark-Gluon Plasma (QGP)~\cite{Harris:1996zx}.  
At the Relativistic Heavy Ion Collider (RHIC) the energy density 
is well in excess of the critical energy density that is expected 
for this transition~\cite{Adcox:2004mh}.
One of the most intriguing results observed at RHIC so far is the
suppression of hadrons with high transverse momentum (\pperp) in central (head-on)
Au+Au collisions. 
The hadron yield at high \pperp is a factor of 5 less than expected from $p+p$ 
collisions scaled by the number of corresponding nucleon-nucleon collisions~\cite{Adcox:2001jp}.
Such suppression was predicted as an effect of 
parton energy loss in the medium generated in the 
collisions~\cite{Bjorken:1982tu,Baier:2000mf}. 
A control experiment of $d$+Au collisions, where no medium is produced 
in the final state of the collision, showed no indication of
hadron suppression at mid-rapidity~\cite{Adler:2003ii}, ruling out strong
initial-state effects (final-state energy loss in the cold nucleus is
generally expected to be small) as the cause for 
the suppression in Au+Au. 
For a better understanding of the medium effects at work in Au+Au,
however, it is crucial to explore the exact role initial-state
effects play in the modification of high-\pperp particle production at RHIC.

Initial-state nuclear effects include the Cronin effect, shadowing, 
and gluon saturation. The Cronin effect~\cite{Cronin:1974zm}, 
an enhancement of the particle yield at intermediate \pperp, 
is usually attributed to multiple soft parton scatterings 
before a hard interaction of the parton ($p_\mathrm{T}$ broadening). 
The shadowing of the structure function~\cite{Arneodo:1992wf}
modifies the particle yield depending on the parton momentum fraction, $x_\mathrm{Bj}$, 
probed in the partonic scattering.
An alternative model of the initial state of a nucleus 
is the gluon saturation or color glass condensate (CGC)
in which the gluon population at low $x_\mathrm{Bj}$ 
is limited 
by non-linear gluon-gluon dynamics.
In this picture, particle
production at moderate \pperp originally was predicted to be
suppressed in central $d$+Au collisions at 
RHIC~\cite{Kharzeev:2002pc}.  In recent CGC models, a Cronin enhancement
can also be reproduced with a suitable choice of initial-state parameters~\cite{Kharzeev:2003wz}.

One established way to test the contribution of different initial- and final-state nuclear
effects is the study of the centrality dependence of particle
production at high \pperp. Initial state and medium effects 
are strongest in central collisions.
In Au+Au collisions a strong dependence of
the suppression of high-\pperp hadrons on the centrality of the
collision has been 
observed~\cite{Adler:2003qi,Adler:2003au,Adams:2003kv}; the suppression weakens
going to peripheral collisions and finally disappears.
This can be compared to the centrality dependence of (initial-state) 
hadron production in $d$+Au. 
The yield of non-identified charged hadrons in $d+$Au collisions 
with $p_\mathrm{T} < 6$~GeV/$c$ was found to be increasingly enhanced going from 
peripheral to central collisions~\cite{ppg041}, mainly attributed to 
the influence of (anti)protons~\cite{Adler:2006xd}.
At high \pperp the baryon contribution to the yield of unidentified
charged hadrons is expected to become small and instead the yield is 
dominated by charged pions~\cite{Adcox:2004mh}. 
All this sparks paramount interest in the centrality dependence of
neutral pion ($\pi^0$) production especially as it
can be measured up to very high \pperp
where particle production is truly perturbative.
Furthermore, the high-\pperp measurement of an additional
identified particle like the eta meson ($\eta$),
with four times the mass of the pion,
may shed light on the question to what extent
the particle-species dependence of the suppression (enhancement) 
observed in Au+Au ($d$+Au) depends on the number of 
constituent quarks rather than on the mass
of the particle~\cite{Adler:2004hv,Hwa:2004zd}.  

%

In this Letter we present measurements by the PHENIX experiment~\cite{nim_phenix} 
on the production of $\pi^0$ and $\eta$ in $p+p$ and 
$d$+Au collisions at $\sqrt{s_{\scriptscriptstyle NN}}$ = 200~GeV. 
The data provides the first measurement of neutral mesons in $d$+Au collisions 
at mid-rapidity  as a function of the centrality of the collision.
The $\pi^0$ measurements described in this paper are similar to the analysis 
of minimum bias $d$+Au data in~\cite{Adler:2003ii} but are based on an improved 
data set that allows the study of the particle production for different selections 
of the centrality of the collision.

%

$\pi^0$ and $\eta$ are measured by the PHENIX electromagnetic calorimeter (EMCal)
via the $\pi^0 \rightarrow \gamma\gamma$ and $\eta \rightarrow \gamma\gamma$ decay. 
The EMCal consists of six lead scintillator (PbSc) and two lead glass (PbGl) sectors, 
each located at a radial distance of $\sim$ 5 m from the beam axis. 
The detector covers a pseudorapidity range of $|\eta| \leq 0.35$ and an azimuthal angle of 
$\Delta \phi = \pi$. 
The EMCal granularity is 
$\Delta\eta \times \Delta\phi \approx 0.011 \times 0.011$ for the PbSc and
0.008 $\times$ 0.008 for the PbGl.
The data sets from PbSc and PbGl are analyzed separately 
and combined for the final results.
The energy calibration for the EMCal is obtained from beam tests, cosmic rays, 
and minimum ionizing energy peaks of charged hadrons.
In a recent improvement of the calibration, the EMCal is calibrated by the invariant mass 
distribution of neutral pions for each of the 24768 readout channels separately.
The uncertainty on the energy scale is $1.2 \%$.

The data used in this analysis was recorded in 2002-2003 (RHIC Run-3) under two different
trigger conditions. 25.2 $\times 10^{6}$ and  58.3 $\times 10^{6}$ minimum bias events 
were analyzed
for $p+p$ and $d$+Au collisions, respectively. Minimum bias (MB) events 
are triggered by the Beam-Beam Counters (BBC)~\cite{nim_phenix} ($|\eta|$ = 3.0--3.9)
and require a vertex position along the beam axis within $|z| <$~30 cm.
The minimum bias trigger accepts $(88 \pm 4)\%$ of all inelastic $d$+Au collisions
that satisfy the vertex condition. 
This corresponds to 1.99 b $\pm 5.2 \%$, the measured fraction of the total $d$+Au
inelastic cross section, determined
using photo-dissociation of the deuteron \cite{White:2005kp}.
In $p+p$ this trigger measures 23.0 mb $\pm 9.7 \%$
of the $p+p$ inelastic cross section.
The measured particle yields are corrected for the $p+p$ MB trigger 
bias~\cite{Adler:2003pb}: 
the MB trigger measures only $(79\pm2)\%$ of high-\pperp particles.
In $d$+Au collisions this fraction varies from 85\% to 100\% from peripheral 
to central collisions;
here the uncertainty is $\sim3\%$.
The second data sample was collected with a high-\pperp photon trigger in the EMCal
in addition to the MB trigger requirement in order to extend the measurement to higher \pperp.
This trigger requires a photon of $p_\mathrm{T} > 1.4 (1.4)$ and $p_\mathrm{T} > 2.5 (3.5)$
for PbSc (PbGl) and for $p+p$ and $d$+Au collisions, respectively. 
We analyzed 45.1 $\times 10^{6}$ (19.5 $\times 10^{6}$) 
events in $p+p$ ($d$+Au) under this trigger condition. 
The sampled integrated luminosity was 216 nb$^{-1}$ for $p+p$ and 1.5 nb$^{-1}$
for $d$+Au. (In $d$+Au that corresponds to an integrated nucleon-nucleon luminosity of 590 nb$^{-1}$).

The division of $d$+Au collisions in different centrality classes is
based on the charge deposited in the backward BBC ($ -3.9 < \eta < -3.0$), 
i. e. in the Au beam direction. For each centrality class the corresponding 
average nuclear overlap function $\langle T_{AB} \rangle$ is calculated
using a Glauber Monte Carlo model and simulations of the BBC, taking into 
account its limited efficiency for peripheral collisions. For the four 
centrality classes ($0-20\%$, $20-40\%$, $40-60\%$ and $60-88\%$)
used in this analysis, $\langle T_{AB} \rangle$ translates into
an average number of nucleon-nucleon collisions per
A+B collision, $\langle N_{\mathrm{coll}}\rangle =
\sigma_{\mathrm{inel}}^{pp} \times \langle
T_{AB}\rangle$, of $(15.4\pm 1.0)$, $(10.6\pm 0.7)$,
$(7.0\pm 0.6)$ and $(3.1\pm 0.3)$, respectively.



%

Photon candidates in the EMCal are selected by applying particle
identification (PID) cuts based on the shower profile in the detector.  
To determine the yields of $\pi^0$ and $\eta$, the invariant mass of 
all photon pairs with an energy asymmetry $|E_1 - E_2|/(E_1+E_2) < 0.7$ 
in a given \pperp bin is calculated. After subtraction of the combinatorial
background the invariant mass distribution is integrated around the
particle mass peak~\cite{Adler:2003qi}; the integration window reflects 
thereby the \pperp dependence of the mass peak position and width.
The combinatorial background is determined by
pairing photons from different events
with similar centrality (for $d$+Au) and vertex.
In this analysis the signal-to-background ratio for high-\pperp $\pi^0$
is about 25 and 13 at \pperp=4~GeV/$c$ in $p+p$ and central $d$+Au collisions, respectively.  
It decreases to 7 and 2 at \pperp=2~GeV/$c$.
For $\eta$, this ratio is about 2 at \pperp=8~GeV/$c$, decreasing to 0.3 ($p+p$) and 0.2 
(central $d$+Au) at \pperp=3~GeV/$c$.
The raw spectra are corrected for trigger efficiency,
acceptance, and reconstruction efficiency. This includes dead areas, 
the influence of energy resolution, analysis cuts, the peak extraction window,
and photon conversion.
The corrections are determined using Monte Carlo simulations. 
Due to the fine granularity of the calorimeter, occupancy effects are negligible.
Furthermore, the $\pi^0$ spectra are corrected at $p_\mathrm{T}>$10~GeV/$c$ (15~GeV/$c$)
for two-photon merging effects in the PbSc (PbGl), studied in Monte Carlo 
simulations and confirmed with test beam data.
Finally a correction in the $\pi^0$ and $\eta$ yields to account for the true 
mean value of each \pperp bin is applied to the steeply falling spectra.
For $p_\mathrm{T} < 3.5$~$(3.0)$~GeV/$c$ the $p+p$ $\pi^0$ ($\eta$) spectrum is calculated
from the minimum bias data sample, above this threshold the high-\pperp
triggered sample is used. In $d$+Au, this transition is made at $p_\mathrm{T} = 4.5$ and $3.5$~GeV/$c$
for $\pi^0$ and $\eta$, respectively.

%

\begin{table}[b]
\caption{\label{tab:syserr}
Main systematic uncertainties in $\%$ on $\pi^0$ and $\eta$ spectra.
The uncertainties are given for PbSc (PbGl). The normalization uncertainties
of $9.7\%$ for the $p+p$ and $5.2\%$ for the $d$+Au cross section
as well as the MB-trigger-bias uncertainty of $\sim3\%$ for the 
centrality-selected yields are not listed.
}
\begin{ruledtabular}
\begin{tabular}{lcccc}
                          meson & $\pi^0$  & $\pi^0$ & $\eta$  & $\eta$  \\

\pperp (GeV/$c$)          &  2        & 15         & 3        & 10 \\
\hline
a) peak extraction        &  2.7(2.7) &  2.0(2.0)  & 14(14)   & 6.0(6.0) \\

b) geom. accept.          &  3.5(3.5) &  3.5(3.5)  & 4.5(4.5) & 4.5(4.5) \\

c) $\pi^0$ reconstr. eff. &  0.7(0.7) &  4.0(4.0)  & 0.7(0.7) & 3.6(3.6) \\

d) energy scale           &  5.0(5.0) & 11.4(11.4) & 5.0(5.0) & 9.4(9.4)  \\

e) merging corr.          &  -        &  5.9(2.1)  & -        & - \\

\hline
Total                     &  6.7(6.7) & 17.0(12.9) & 15.5(15.5) & 12.6(12.6) \\



 \end{tabular}
 \end{ruledtabular}
 \end{table}

The main contributions to the systematic uncertainty on the $p+p$ and $d$+Au spectra
are given in Table~\ref{tab:syserr} for $\pi^0$ and $\eta$.
Most uncertainties are identical for $p+p$ and $d$+Au, only the uncertainty on
the peak extraction is slightly larger in $d$+Au.
Category (d) includes uncertainties on the EMCal global energy scale and non-linearity. 
The uncertainties in (d) and (e) are partially correlated. 
All others are uncorrelated. 



%

 \begin{figure}
 \includegraphics[width=1.0\linewidth]{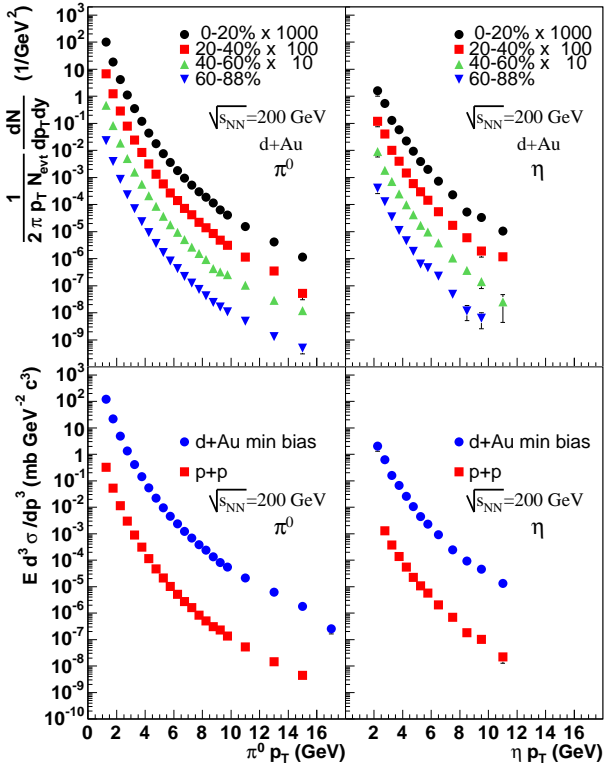}
 \caption{\label{fig:pt_spectra}
 Top: invariant yields at mid-rapidity for $\pi^0$ (left) and $\eta$
 (right) in $d$+Au collisions as a function of \pperp for different selections 
 of the centrality of the collision.
 Bottom: invariant cross section at mid-rapidity for $\pi^0$ (left) and $\eta$
 (right) in $p+p$ and $d$+Au collisions as a function of \pperp.
 }
 \end{figure}

The fully corrected \pperp distributions of $\pi^0$ and $\eta$
are shown in Fig.~\ref{fig:pt_spectra}. 
The top panels show the invariant yield in $d$+Au collisions for four
centrality bins scaled for clarity by the factors indicated.  The
bottom panels show the invariant cross section in $p+p$ and $d$+Au
collisions.
The improved dataset 
allows the study of $\pi^0$ ($\eta$) production up to 
18 (12)~GeV/$c$, the highest \pperp values measured
for identified particles in $p(d)$+A collisions.  
For the first time, the invariant 
cross section for $\pi^0$ and $\eta$
in $d$+Au collisions has been measured at this energy.
The $\pi^0$ result in $p+p$ agrees with the previous measurement 
at $\sqrt{s_{\scriptscriptstyle NN}}$ = 200~GeV~\cite{Adler:2003pb}
within statistical uncertainties, and confirms the
agreement with pQCD within the uncertainty of the calculation.
Therefore the $p+p$ cross section can be used as a well-understood
reference for the production in $d$+A and Au+Au collisions.

%

 \begin{figure}[t]
 \includegraphics[width=1.0\linewidth]{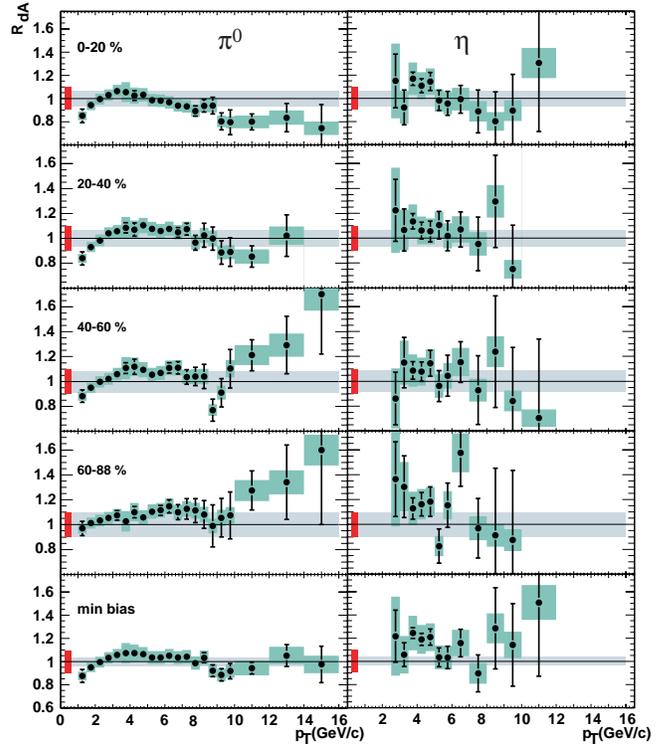}
 \caption{\label{fig:RdA}
 Nuclear modification factor $R_{\rm{dA}}$ for $\pi^0$ and $\eta$ in different
 centrality selections and min. bias data. The bands around the data
 points show systematic errors which can vary with \pperp. The shaded 
 bands around unity indicate the $\langle T_{AB}\rangle$ uncertainty
 and the small bands on the left side of the data points
 indicate the normalization uncertainty due to the $p+p$ reference.
 }
 \end{figure}

To quantify nuclear medium effects at high $p_{T}$ it is customary to
use the {\it nuclear modification factor} which is given by the ratio 
of the $d$+Au yield to the {\it p+p} cross section~\cite{Adler:2003qi}
scaled by $\langle T_{AB}\rangle$:
\begin{equation}
R_{AB}(p_{T})\,=\,\frac{d^2N^{\pi^0}_{AB}/dy dp_{T}}{\langle T_{AB}\rangle\,\times\, d^2\sigma^{\pi^0}_{pp}/dy dp_{T}}.
\label{eq:R_AA}
\end{equation}
The average nuclear overlap function $\langle T_{AB}\rangle$,
averaged over the respective impact parameter range, is
determined solely by the density distribution of the nucleons in the nuclei A and B
and the impact parameter.

Figure \ref{fig:RdA} shows the nuclear modification factor $R_{dA}(p_\mathrm{T})$
for $\pi^0$ and $\eta$ in $d$+Au collisions at $\sqrt{s_{\scriptscriptstyle NN}}$=200~GeV
for four different centrality selections and for minimum bias events.
As the $p+p$ and $d$+Au measurements
were both made in the same year, many of the systematic errors associated
with detector performance were nearly identical and the corresponding
systematic errors in the comparison are negligible.
Within systematic errors $R_{dA}(p_\mathrm{T})$ for $\pi^0$ and $\eta$
is $\approx 1$ in all centrality bins, and only a weak \pperp dependence can be seen.
In order to check the absolute normalization systematics we can also
calculate $R_{dA}(p_\mathrm{T})$ using the inelastic cross section measured through 
photo-dissociation of the deuteron.
This constitutes an important cross check. It replaces the systematic
uncertainties of the BBC efficiency and $\langle T_{AB}\rangle$, which are determined
by model calculations, by the uncertainty of the cross section measurement
of similar size.
The resulting $R_{dA}(p_\mathrm{T})$ is 9.8 \%
larger than that obtained from the minimum bias yield, consistent within 1.5 $\sigma$.

Though very different in mass, $\eta$ and $\pi^0$ show a similar, weak centrality
dependence of $R_{dA}(p_\mathrm{T})$ over the measured \pperp range. 
These results do not show the significant enhancement seen for protons
where the proton $R_{AA}$ is substantially larger than that of pions in the
intermediate \pperp (2 GeV/$c <$ \pperp $< 4 $ GeV/$c$) region \cite{Adler:2006xd}.
The $\pi^0$ data exhibit small shape variations with centrality that may be
due to initial-state effects including shadowing and multiple scattering.  
Possible Cronin enhancements in the intermediate \pperp region due to
initial-state multiple scattering or anti-shadowing are not more than $10\%$
around 4 GeV/$c$. At low 
\pperp (\pperp $< 3$ GeV/$c$) the drop towards smaller $R_{dA}$ is
consistent with analogous measurements for charged pions \cite{Adler:2006xd} and is
usually attributed to a change to a regime of soft physics ($N_{\mathrm part}$ scaling)
at the smallest \pperp values. 
At the largest \pperp values measured (\pperp$>9$ GeV/$c$)
the most central $\pi^0$ 
result hints at a small suppression, though this is only a $\sim 1.7$ sigma effect.

%
In conclusion, we have presented the first study of the centrality dependence of
$\pi^0$ and $\eta$ production at mid-rapidity in $d$+Au collisions 
at $\sqrt{s_{\scriptscriptstyle NN}}$ = 200~GeV.
Transverse momentum spectra up to $p_\mathrm{T} =  18$ and 12~GeV/$c$ have
been measured for $\pi^0$ and $\eta$, respectively.
The invariant yield per nucleon-nucleon collision is compared to that in $p+p$ collisions
measured at the same $\sqrt{s_{\scriptscriptstyle NN}}$.  The strong
suppression observed for $\pi^0$ production at high \pperp in central 
Au-Au collisions cannot be seen for $d$+Au in any centrality:
Within systematic errors $R_{dA}(p_\mathrm{T})$ is $\approx 1$ in all centrality bins.
A weak centrality dependence of the shape of $R_{dA}$ versus \pperp is seen,
presumably due to initial-state effects.  
A possible Cronin enhancement is substantially smaller than
the $R_{dA} \gsim 1.9$ that corresonds to results from lower energy measurements 
\cite{Cronin:1974zm,Angelis:1985fk}.
Within systematic errors $R_{dA}$ for $\pi^0$ and $\eta$
agrees well, giving no indication for cold nuclear
matter effects having a mass dependence.
Since nuclear modifications in $d$+Au are small even in the most central collisions
where initial state effects are expected to be largest, 
we conclude that initial state effects in Au+Au must be small as well, so the
large suppression seen in Au+Au must be mostly due to medium effects.


We thank the staff of the Collider-Accelerator and Physics
Departments at BNL for their vital contributions.  
We acknowledge support from 
the Department of Energy and NSF (U.S.A.), 
MEXT and JSPS (Japan), 
CNPq and FAPESP (Brazil), 
NSFC (China), 
IN2P3/CNRS, CEA, and ARMINES (France), 
BMBF, DAAD, and AvH (Germany), 
OTKA (Hungary), 
DAE and DST (India), 
ISF (Israel), 
KRF and KOSEF (Korea), 
RMIST, RAS, and RMAE (Russia), 
VR and KAW (Sweden), 
U.S. CRDF for the FSU, 
US-Hungarian NSF-OTKA-MTA, 
and US-Israel BSF.



\end{document}